\documentclass[11pt,a4paper]{article}

\makeatletter 

\usepackage{color, amsmath, amssymb, wrapfig, url, slashed, cite, ifpdf}
\usepackage{multirow, feynmp}

\renewcommand\citepunct{,\penalty\@M\hskip.13emplus.1emminus.1em\relax} 
\usepackage{enumerate}

\ifpdf                                     
  \usepackage{hyperref}
  \usepackage{graphicx}           
\else                                      
  \usepackage[dvipdfmx]{hyperref,graphicx} 
\fi

\newcommand{\dd}{{\mathrm d}}      

\renewcommand{\Re}{\mathop{\mathrm{Re}}}

\newcommand{\@diff}   [4]{\dfrac{#4#3#1}{#4#2#3}}
\newcommand{\diff}    [2]{\@diff{#1}{#2}{}{\dd}}
\newcommand{\pdiff}   [2]{\@diff{#1}{#2}{}\partial}
\newcommand{\fdiff}   [2]{\@diff{#1}{#2}{}{\delta}}
\newcommand{\ndiffnum}[3]{\@diff{#1}{#2}{^#3}\dd}
\newcommand{\npdiff}  [3]{\@diff{#1}{#2}{^#3}\partial}
\newcommand{\@difftwo}[4]{\dfrac{#4^2#1}{#4#2\,#4#3}}
\newcommand{\difftwo} [3]{\@difftwo{#1}{#2}{#3}{\dd}}
\newcommand{\pdifftwo}[3]{\@difftwo{#1}{#2}{#3}{\partial}}

\newcommand{\un}[1]{{\mathrm{\,#1}}} 
\newcommand{\TeV}{\un{TeV}}
\newcommand{\GeV}{\un{GeV}}
\newcommand{\MeV}{\un{MeV}}
\newcommand{\keV}{\un{keV}}

\def\@xxxEV{\@ifnextchar-{\@xxxEV@minus}{\@xxxEV@plus}}
\def\@xxxEV@plus#1#2{%
  \ifnum{#1=0}{}\else\ifnum{#1=1}{10}\else {10^#1}\fi\fi #2}
\def\@xxxEV@minus#1#2 {10^{-#1}{\rm\,#2}}

\newcommand{\TEV}[1]{\@xxxEV{#1}{\TeV}}
\newcommand{\GEV}[1]{\@xxxEV{#1}{\GeV}}
\newcommand{\MEV}[1]{\@xxxEV{#1}{\MeV}}
\newcommand{\KEV}[1]{\@xxxEV{#1}{\keV}}

\def\EE{\@ifnextchar-{\@@EE}{\@EE}}
\def\@EE#1{\ifnum#1=1\times10\else\times10^{#1}\fi}
\def\@@EE#1#2{\!\times\!10^{-#2}}

\def\T{\@ifnextchar^{\T@u}{\@ifnextchar_{\T@d}{}}}
\def\T@u^#1{{^{#1}}\T}
\def\T@d_#1{{_{#1}}\T}

\newcommand{\s}[1]{_\mathrm{#1}}    

\hypersetup{pdfauthor={Motoi Endo and Sho Iwamoto},pdftitle={Comment on the CMS search for charge--asymmetric production of W-prime boson in ttbar+jet events}}

\newcommand{\dummybibliography}[1]{\relax}
\newcommand{\AFB}{A\s{FB}}
\newcommand{\ttbar}{t\bar t}
\newcommand{\PR}{P\s R}

\newcommand{\gR}{g\s R}

\makeatother 

\begin{document}

\begin{titlepage}

\begin{flushright}
UT--12--25
\end{flushright}

\vskip 3cm
\begin{center}
{\Large \bf
Comment on the CMS search for charge--asymmetric production of $W'$ boson in $\ttbar+\text{jet}$ events
}
\vskip 1.2cm
Motoi Endo, Sho Iwamoto
\vskip 0.9cm

{\it Department of Physics, University of Tokyo,
Tokyo 113--0033, Japan
}

\vskip 3cm

\abstract{
A reanalysis is presented on the CMS result on a search for a $W'$ boson that couples to the top and down quarks. The model is motivated by the Tevatron results on the forward--backward asymmetry of $\ttbar$ pair production. In the evaluation of the theoretical cross section of $pp \to \ttbar j$, the interference effect between the SM and $W'$ amplitudes is shown to be important, though it is ignored in the CMS analysis. The lower mass bound on the $W'$ boson is relaxed from $840\GeV$ to $740\GeV$ at the 95\% C.L. due to the interference effect. The bound is also compared to the top forward-backward asymmetry. 
}
\end{center}
\end{titlepage}

\setcounter{page}{2}

The CDF and D0 experiments measured the forward--backward (FB) asymmetry of the top quark pair production, which is defined as
\begin{align}
 \AFB = \frac{\text{\#events}(\Delta y>0) - \text{\#events}(\Delta y<0)}{\text{\#events}(\Delta y>0) + \text{\#events}(\Delta y<0)},
\end{align}
where $\Delta y = y(t)-y(\bar t)$ is the rapidity difference. The measured inclusive asymmetry is
\begin{align}
 \text{CDF \cite{CDF10807}      : } & \AFB=0.162 \pm 0.047, &
 \text{D0  \cite{Abazov:2011rq} : } & \AFB=0.196 \pm 0.065,
\end{align}
at the parton level, which is about $2.2\sigma$ away from the Standard Model (SM) prediction, $\AFB = 0.087 \pm 0.010$~\cite{Kuhn:2011ri}. Categorized by the rapidity difference, a discrepancy is found in a large $|\Delta y|$ region, whose experimental result is
\begin{equation}
 \AFB(|\Delta y| > 1) = 0.433 \pm 0.193,
\end{equation}
from the CDF \cite{CDF10807}. This is $2.2\sigma$ larger than the SM value, $\AFB(|\Delta y| > 1) = 0.193 \pm 0.015$~\cite{Kuhn:2011ri}. The difference is enhanced in a large invariant mass region of $\ttbar$. The CDF reported \cite{CDF10807}
\begin{equation}
 \AFB(m_{\ttbar} > 450\GeV) = 0.296 \pm 0.067,
\end{equation}
while the SM expectation is $\AFB(m_{\ttbar} > 450\GeV) = 0.128 \pm 0.011$~\cite{Kuhn:2011ri}, leading to $2.5\sigma$ deviation. The inconsistency is also observed in the leptonic channel of the $\ttbar$ decay. The D0 measured the leptonic asymmetry as \cite{abazov:2012bf}
\begin{equation}
 \AFB^l =0.118 \pm 0.032,
\end{equation}
which is compared to the SM, $\AFB^l =0.047 \pm 0.001$, providing $2.2\sigma$ discrepancy. The $W + \text{4-jets}$ background could be a source of these deficits~\cite{Hagiwara:2012td}, whereas the deviations might be a sign of physics beyond the SM.

In order to explain the anomaly, a model with a flavor--changing $W'$ boson was proposed~\cite{Cheung:2009ch}, whose Lagrangian is
\begin{align}
 \mathcal L_{W'} = \gR \left(W'_\mu\bar d\gamma^\mu\PR t+\text{h.c.}\right).
\end{align}
This can contribute to the asymmetry by $\sim 0.1$, which is a size of the current discrepancy. Motivated by this prospect, the CMS collaboration recently reported a result on a search for the $\ttbar+\text{jet}$ event at the center-of-mass energy of $\sqrt{s}=7\TeV$ with the integrated luminosity of $5.0{\rm fb}^{-1}$ \cite{Chatrchyan:2012su}. They used a theoretical cross section at the leading order (LO), referring to Ref.~\cite{Knapen:2011hu}. However, the cross section is evaluated with neglecting the interference between the SM and $W'$ amplitudes. In this note, we show that the interference effect is comparable to the cross section and must be included. The mass bound on $W'$ is found to be relaxed by $\sim 100\GeV$ compared to that obtained in the CMS paper.


\begin{figure}[p]\begin{center}
\begin{fmffile}{feyn_dg}
\begin{fmfgraph*}(100,80)\fmfstraight
\fmfleft{x1,a2,x2,x9,x8,a1,x3}\fmfright{x4,b3,x5,b2,x7,b1,x6}
\fmf{phantom}{x9,xx0,v1,xx1,v2,xx2,xx3,,b2}
\fmffreeze
\fmf{xquark}{a1,v1,v2,b3} \fmf{gluon}{a2,v1} \fmf{photon,lab=$W'$}{v2,v3} \fmf{phantom}{v2,b1}
\fmf{xquark}{b1,v3,b2}
\fmfdot{v2,v3} \fmfv{l=$d$,l.a=180}{a1}\fmfv{l=$g$,l.a=180}{a2}
\fmfv{l=$d$,l.a=0}{b1}\fmfv{l=$\bar t$,l.a=0}{b2}\fmfv{l=$t$,l.a=0}{b3}
\end{fmfgraph*}
\hspace{60pt}
\begin{fmfgraph*}(100,80)\fmfstraight
\fmfleft{x1,a2,x2,x9,x8,a1,x3}\fmfright{x4,b3,x5,b2,x7,b1,x6}
\fmf{phantom,tension=2}{a1,v1}\fmf{phantom,tension=2}{v2,a2}\fmf{phantom,tension=1}{v1,v2}
\fmf{phantom}{b1,v1}\fmf{phantom}{v2,b3}
\fmffreeze
\fmf{xquark}{a1,v1,v2,b3} \fmf{gluon}{a2,v2} \fmf{photon,lab=$W'$}{v1,v3} \fmf{phantom}{v2,b1}
\fmf{xquark}{b1,v3,b2}
\fmfdot{v1,v3} \fmfv{l=$d$,l.a=180}{a1}\fmfv{l=$g$,l.a=180}{a2}
\fmfv{l=$d$,l.a=0}{b1}\fmfv{l=$\bar t$,l.a=0}{b2}\fmfv{l=$t$,l.a=0}{b3}
\end{fmfgraph*}

\begin{fmfgraph*}(100,80)\fmfstraight
\fmfleft{x1,a2,x2,x9,x8,a1,x3}\fmfright{x4,b3,x5,b2,x7,b1,x6}
\fmf{xquark}{a1,v1,b1} \fmf{gluon}{a2,v2} \fmf{phantom}{v2,b3} \fmf{phantom}{a1,v3,b3}
\fmffreeze
\fmf{photon,lab=$W'$}{v1,v3} \fmf{xquark}{b2,v3,v2,b3}
\fmfdot{v1,v3}\fmfv{l=$d$,l.a=180}{a1}\fmfv{l=$g$,l.a=180}{a2}
\fmfv{l=$t$,l.a=0}{b1}\fmfv{l=$\bar t$,l.a=0}{b2}\fmfv{l=$d$,l.a=0}{b3}
\end{fmfgraph*}
\hspace{60pt}
\begin{fmfgraph*}(100,80)\fmfstraight
\fmfleft{x1,a2,x2,x9,x8,a1,x3}\fmfright{x4,b3,x5,b2,x7,b1,x6}
\fmf{xquark}{a1,v1,b1} \fmf{gluon}{a2,v2} \fmf{phantom}{v2,b3} \fmf{phantom}{a1,v3,b3}
\fmffreeze
\fmf{photon,lab=$W'$}{v1,v3} \fmf{xquark}{b2,v3,v2,b3}
\fmfdot{v1,v3} \fmfv{l=$d$,l.a=180}{a1}\fmfv{l=$g$,l.a=180}{a2}
\fmfv{l=$t$,l.a=0}{b1}\fmfv{l=$d$,l.a=0}{b2}\fmfv{l=$\bar t$,l.a=0}{b3}
\end{fmfgraph*}

\vspace{20pt}

\begin{fmfgraph*}(100,80)\fmfstraight
\fmfleft{x1,x8,x9,a2,x2,a1,x3}\fmfright{x4,b3,x7,b2,x5,b1,x6}
\fmf{xquark}{a1,v1,b1}
\fmf{xquark}{a2,v2,b2}
\fmf{phantom}{x9,xx1,v3,x7}
\fmf{gluon,tension=0}{v3,b3}
\fmffreeze
\fmf{photon,lab=$W'$}{v1,v2}
\fmfdot{v1,v2} \fmfv{l=$d$,l.a=180}{a1}\fmfv{l=$\bar d$,l.a=180}{a2}
\fmfv{l=$t$,l.a=0}{b1}\fmfv{l=$\bar t$,l.a=0}{b2}\fmfv{l=$g$,l.a=0}{b3}
\end{fmfgraph*}
\end{fmffile}\end{center}
\vspace{-20pt}
\caption{The $W'$ boson contribution to the events with $\ttbar+\text{jet}$ final state. The first four diagrams generate $\ttbar+d$ events, and the last contributes to $\ttbar + g$ events.
Only the first two diagrams have the $s$-channel $W'$ production.}
\label{fig:dg}
\end{figure}
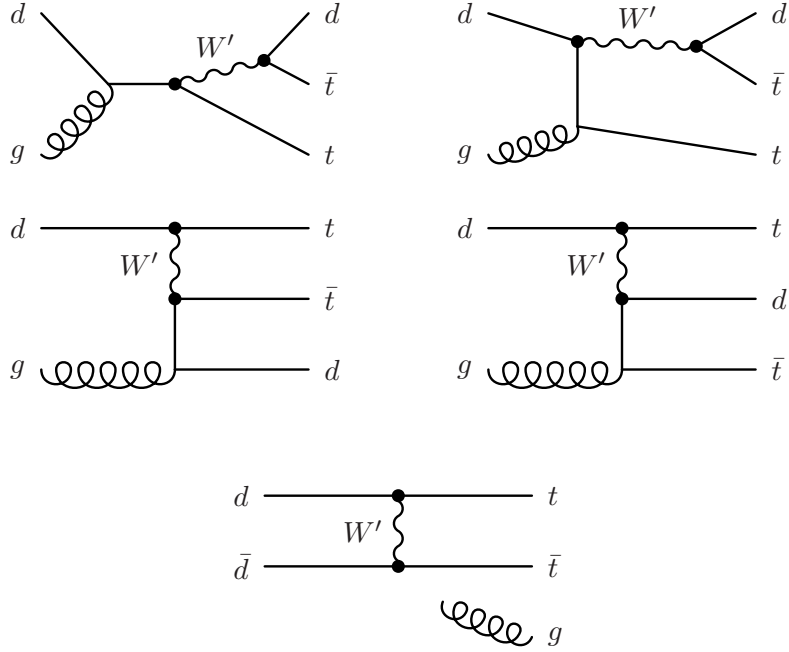

\begin{figure}[p]
\begin{center}\begin{fmffile}{feyn_qcd}
\begin{fmfgraph*}(100,80)\fmfstraight
\fmfleft{x1,x9,x8,a2,x2,a1,x3}\fmfright{x4,b3,x7,b2,x5,b1,x6}
\fmf{xquark}{a1,v1,a2}
\fmf{xquark}{b1,v2,b2}
\fmf{phantom}{a2,xx1,v3,b2}
\fmf{gluon,tension=0}{v3,b3}
\fmf{gluon,tension=0.7}{v1,v2}
\fmfv{l=$d$,l.a=180}{a1}\fmfv{l=$\bar d$,l.a=180}{a2}
\fmfv{l=$t$,l.a=0}{b1}\fmfv{l=$\bar t$,l.a=0}{b2}\fmfv{l=$g$,l.a=0}{b3}
\end{fmfgraph*}
\end{fmffile}\end{center}
\caption{Part of the SM contribution to the $\ttbar + g$ events. This interferes with the last diagram in Fig.~\ref{fig:dg}.}
\label{fig:qcd}
\end{figure}
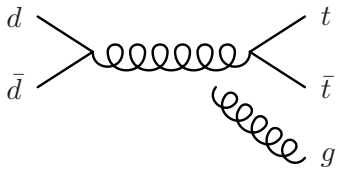

The CMS searched for the $pp \to \ttbar j$ event.
The $W'$ boson contributes to the processes $gd(\bar d)\to\ttbar d(\bar d)$ and $d\bar d\to\ttbar g$ as shown in Fig.~\ref{fig:dg}. Since both processes have also the SM contributions, the total cross section is represented as
\begin{equation}
 \left|\mathcal M\right|^2
= \left|\mathcal M\s{SM} + \mathcal M_{W'}\right|^2 
= \left|\mathcal M\s{SM}\right|^2 + \left|\mathcal M_{W'}\right|^2 + \mathcal E\s{interference}
\end{equation}
for each process, where $\mathcal E\s{interference}$ denotes the interference term, $\mathcal E\s{interference} \equiv 2\Re(\mathcal M\s{SM}\mathcal M_{W'}^*)$. It is emphasized that the deviation of the cross section from the SM is calculated as {\it the sum} of $\left|\mathcal M_{W'}\right|^2$ and $\mathcal E\s{interference}$. The process $pp \to \ttbar g$ has a large contribution from the SM through the diagram Fig.~\ref{fig:qcd}, and thus, $\mathcal E\s{interference}$ for the process is as large as the $W'$ contribution, $\left|\mathcal M_{W'}\right|^2$.

The cross sections with and without the interference are calculated with {\tt MadGraph\,5}~\cite{MadGraph5}, using the model file generated by ourselves with {\tt FeynRules\,1.6}~\cite{FeynRules}. The simulation is based on the LO calculation, and {\tt CTEQ6L1}~\cite{PDFCTEQ6} is used as the parton distribution function.

\setcounter{footnote}{-1} 
\begin{figure}[t]\begin{center}
\includegraphics[width=300pt]{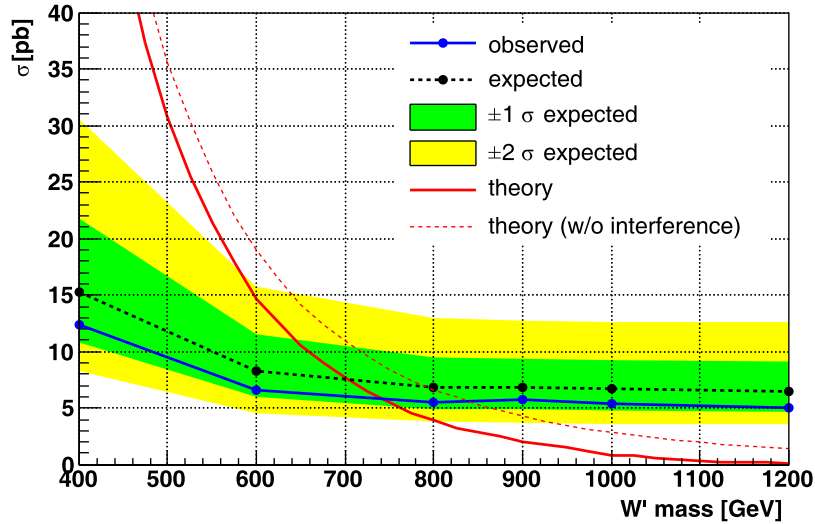} 
 \caption[The 95\% C.L. expected and observed limits on $W'$ production for $\gR=2$ as a function of the $W'$ boson mass, which is reported by the CMS collaboration \cite{Chatrchyan:2012su}, together with the theoretical production section with and without the interference effect.]
{The 95\% C.L. expected and observed limits on $W'$ production for $\gR=2$ as a function of the $W'$ boson mass, which is reported by the CMS collaboration \cite{Chatrchyan:2012su}%
\footnotemark
, together with the theoretical production section with and without the interference effect.}
 \label{fig:cmsplot}
\end{center}\end{figure}
\footnotetext{The numerical values to write this plot (except the theory lines) are obtained from an web page of the CMS collaboration, \url{http://twiki.cern.ch/twiki/bin/view/CMSPublic/PhysicsResultsEXO11056}.}

In Fig.~\ref{fig:cmsplot}, the theoretical cross section is compared to the exclusion bound by the CMS \cite{Chatrchyan:2012su}. We reproduced the referred cross section in the CMS paper by ignoring the interference effect, which is shown by the red-dotted line in Fig.~\ref{fig:cmsplot}. Based on the analysis, they put a limit of $m_{W'} > 840\GeV$ at the 95\% C.L. if the $W'$ coupling is fixed to be $\gR=2$.
However, including the interference effect, the cross section reduces as shown in Fig.~\ref{fig:cmsplot} with the red-solid line. 
The theoretical cross section decreases by $\sim 40$\% for $m_{W'} = 800\GeV$.
We found that, due to the interference effect, the lower bound is corrected as $m_{W'} > 740\GeV$ at the 95\% C.L. for $\gR=2$.

\begin{table}[t]
 \begin{center}
\catcode`?=\active \def?{\phantom{0}} 
 \begin{tabular}[t]{|c|c|c|c|c|c|c|c|}\hline
\multirow{2}{*}{process} & \multirow{2}{*}{SM}
&\multicolumn{3}{|c|}{$M_{W'}=600\GeV, \gR=0.6$}
  & \multicolumn{3}{|c|}{$M_{W'}=600\GeV, \gR=2.0$}\\\cline{3-8}
& &  $W'$ only & interfer. &  total &  $W'$ only & interfer.     &  total \\\hline
$gg\to t\bar t g$                  & $50.6$ &  ---  &   ---  & $50.6$ &  ---  &   ---  &$?50.6$ \\
$gu(\bar u)\to t\bar t u(\bar u)$  & $22.4$ &  ---  &   ---  & $22.5$ &  ---  &   ---  &$?22.5$ \\
$gd(\bar d)\to t\bar t d(\bar d)$  & $10.8$ & $0.7$ & $-0.2$ & $11.3$ &$10.6$ & $+0.2$ &$?21.6$ \\
$d\bar d\to t\bar t g$             & $?2.6$ & $0.1$ & $-0.3$ & $?2.4$ &$?8.4$ & $-3.8$ &$??7.2$ \\
 the others                        & $?7.4$ &  ---  &   ---  & $?7.4$ &  ---  &   ---  &$??7.4$ \\
\hline 
  all processes                    & $93.9$ & $0.8$ & $-0.5$ & $94.2$ &$19.0$ & $-3.6$ &$109.3$ \\
\hline
 \end{tabular}
 \end{center}
 \caption{Cross sections (in units of pb) of each processes at two benchmark points calculated with {\tt MadGraph\,5}. ``The others'' includes $u\bar u\to t\bar t g$, and the processes with $s$- and $c$-quarks. Statistical uncertainties are tiny enough, while theoretical uncertainties are not considered.}
 \label{tab:cs}
\end{table}

Let us investigate the interference effect. In Tab.~\ref{tab:cs}, the cross sections are shown for two benchmark points of $(m_{W'}, \gR)$.
For the process $gd\to\ttbar d$, the interference term is sub-leading compared to the total $W'$ contribution. Especially when $\gR$ is small, the $s$-channel production of the $W'$ boson dominates. 
On the other hand, since the process $d\bar d\to \ttbar g$ has a sizable SM contribution, the interference effect is enhanced.
Since the effect works destructive, focusing on the $s$-channel $W'$ production would be a good choice to increase the signal-to-background ratio, which can be done with a cut on the invariant mass of the jet and the $\bar t$-quark. With the cut the first two diagrams of Fig.~\ref{fig:dg} mainly contributes, and destructive interference in $d\bar d\to\ttbar g$ events would be ignorable.

\begin{figure}[t]\begin{center}
\includegraphics[width=300pt]{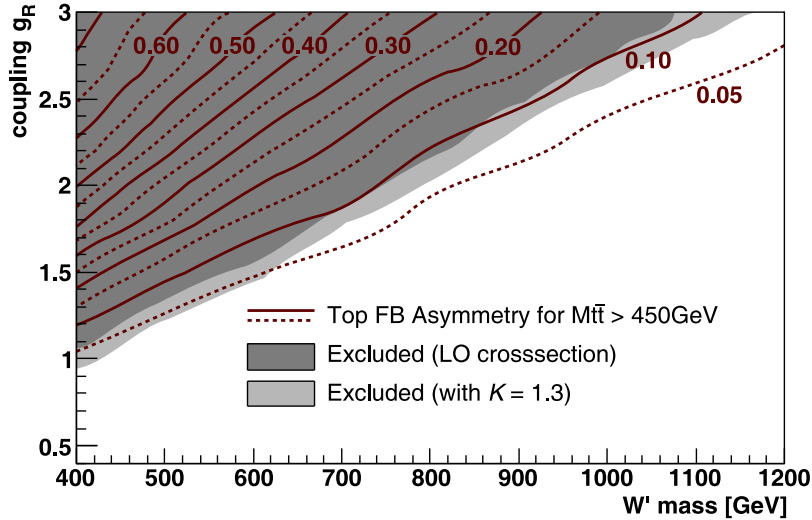}
 \caption{$\AFB(m_{\ttbar} > 450\GeV)$ in the $W'$ model together with the 95\% C.L. observed limits on $W'$ production cross section. }
 \label{fig:twodplot}
\end{center}\end{figure}

The limit on the cross section in Fig.~\ref{fig:cmsplot} is converted to the bound on the $m_{W'}$--$\gR$ plane.~\footnote{A similar plot is found in Ref.~\cite{Duffty:2012zz}, where the ATLAS search for the $\ttbar + j$ events at the accumulated data of $0.7{\rm fb}^{-1}$ \cite{ATLAS2011100} is used. The latest CMS result \cite{Chatrchyan:2012su} is found to provide a more severe limit on the parameter space. }
Taking account of the interference effect, the dark gray region in Fig.~\ref{fig:twodplot} is excluded by the CMS search for $\ttbar$+jet.
The QCD correction can change the cross section. If the $K$-factor of 1.3 is included according to the argument in Ref.~\cite{Knapen:2011hu}, the excluded region becomes wider, which is shown by the light gray region in Fig.~\ref{fig:twodplot}.
Note that we employed an approximation to draw the excluded region that the acceptance does not depend on the coupling $\gR$; the CMS result, i.e. the blue solid line in Fig.~\ref{fig:cmsplot}, is used as the limit on the cross section.

The CMS bound is compared to the model prediction of the top FB asymmetry with $m_{\ttbar} > 450\GeV$, which is described as the contours in Fig.~\ref{fig:twodplot}. The asymmetry is evaluated with {\tt MadGraph\,5} at the LO level. 
Since the asymmetry emerges at the NLO level of the QCD in the SM, $\AFB$ in Fig.~\ref{fig:twodplot} originates in the $W'$ contribution. It is found that $\AFB(m_{\ttbar} > 450\GeV)$ is limited to be less than $0.1$, which looks insufficient to explain the Tevatron results.

Finally, let us comment on the acceptance and the QCD corrections.
The acceptance depends on the jet distributions in the detectors.
The interference effect changes the relative size of each contribution, which could affect the acceptance.
According to the CMS analysis \cite{Chatrchyan:2012su}, this seems more important for heavier $W'$.
Moreover, the top FB asymmetry at the parton level also depends on the acceptance at the CDF and D0 detectors~\cite{Gresham:2011pa,Jung:2011zv,Grinstein:2011dz}.
Since the corrections are not included in this note, the full detector analysis is required.
Next, the QCD correction is not considered for the evaluation of the top FB asymmetry.
The NLO correction to the $W'$ contribution can increase the asymmetry by about 10\%~\cite{Yan:2011tf}.
Nonetheless, the CMS bound on the $\ttbar+\text{jet}$ event provides a crucial constraint on $\AFB$ in the flavor--changing $W'$ model.
Also as discussed above, focusing on $s$-channel $W'$ production is considered to feature the signal.
The authors are eager for the CMS to update the analysis with the interference effect between the SM and $W'$ amplitudes.

\subparagraph{Acknowledgment}
The work by M.E. was supported by Grand-in-Aid for Scientific research from the Ministry of
Education, Science, Sports, and Culture (MEXT), Japan, No.~23740172.
The work by S.I. was supported by Japan Society for the Promotion of Science (JSPS) Grant-in-Aid for JSPS Fellows.


\dummybibliography{ExpResLHC,TopAFB}
\bibliography{}

\begingroup\raggedright\begin{thebibliography}{10}

\bibitem{CDF10807}
{\bfseries CDF} Collaboration,
 CDF Note 10807.

\bibitem{Abazov:2011rq}
{\bfseries D0} Collaboration,
 \href{http://dx.doi.org/10.1103/PhysRevD.84.112005}{Phys.Rev. {\bfseries D84}
  (2011) 112005}
{\ttfamily [\href{http://arxiv.org/abs/1107.4995}{arXiv:1107.4995}]}.

\bibitem{Kuhn:2011ri}
J.~H.~Kuhn and G.~Rodrigo,
 \href{http://dx.doi.org/10.1007/JHEP01(2012)063}{JHEP {\bfseries 1201} (2012)
  063}
{\ttfamily [\href{http://arxiv.org/abs/1109.6830}{arXiv:1109.6830}]}.

\bibitem{abazov:2012bf}
{\bfseries D0} Collaboration,
{\ttfamily \href{http://arxiv.org/abs/1207.0364}{arXiv:1207.0364}}.

\bibitem{Hagiwara:2012td}
K.~Hagiwara, J.~Kanzaki, and Y.~Takaesu,
{\ttfamily \href{http://arxiv.org/abs/1205.5173}{arXiv:1205.5173}}.

\bibitem{Cheung:2009ch}
K.~Cheung, W.-Y.~Keung, and T.-C.~Yuan,
 \href{http://dx.doi.org/10.1016/j.physletb.2009.11.015}{Phys. Lett. {\bfseries
  B682} (2009) 287--290}
{\ttfamily [\href{http://arxiv.org/abs/0908.2589}{arXiv:0908.2589}]}.

\bibitem{Chatrchyan:2012su}
{\bfseries CMS} Collaboration,
{\ttfamily \href{http://arxiv.org/abs/1206.3921}{arXiv:1206.3921}}.

\bibitem{Knapen:2011hu}
S.~Knapen, Y.~Zhao, and M.~J.~Strassler,
{\ttfamily \href{http://arxiv.org/abs/1111.5857}{arXiv:1111.5857}}.

\bibitem{MadGraph5}
J.~Alwall, M.~Herquet, F.~Maltoni, O.~Mattelaer, and T.~Stelzer,
 \href{http://dx.doi.org/10.1007/JHEP06(2011)128}{JHEP {\bfseries 06} (2011)
  128}
{\ttfamily [\href{http://arxiv.org/abs/1106.0522}{arXiv:1106.0522}]}.

\bibitem{FeynRules}
N.~D.~Christensen and C.~Duhr,
 \href{http://dx.doi.org/10.1016/j.cpc.2009.02.018}{Comput. Phys. Commun.
  {\bfseries 180} (2009) 1614--1641}
{\ttfamily [\href{http://arxiv.org/abs/0806.4194}{arXiv:0806.4194}]}.

\bibitem{PDFCTEQ6}
J.~Pumplin {\em et al.},
 \href{http://dx.doi.org/10.1088/1126-6708/2002/07/012}{JHEP {\bfseries 07}
  (2002) 012}
{\ttfamily [\href{http://arxiv.org/abs/hep-ph/0201195}{hep-ph/0201195}]}.

\bibitem{Duffty:2012zz}
D.~Duffty, Z.~Sullivan, and H.~Zhang,
 \href{http://dx.doi.org/10.1103/PhysRevD.85.094027}{Phys.Rev. {\bfseries D85}
  (2012) 094027}
{\ttfamily [\href{http://arxiv.org/abs/1203.4489}{arXiv:1203.4489}]}.

\bibitem{ATLAS2011100}
{\bfseries ATLAS} Collaboration,
 \href{http://cdsweb.cern.ch/record/1369215/}{ATLAS-CONF-2011-100}.

\bibitem{Gresham:2011pa}
M.~I.~Gresham, I.-W.~Kim, and K.~M.~Zurek,
 \href{http://dx.doi.org/10.1103/PhysRevD.83.114027}{Phys. Rev. {\bfseries D83}
  (2011) 114027}
{\ttfamily [\href{http://arxiv.org/abs/1103.3501}{arXiv:1103.3501}]}.

\bibitem{Jung:2011zv}
S.~Jung, A.~Pierce, and J.~D.~Wells,
 \href{http://dx.doi.org/10.1103/PhysRevD.83.114039}{Phys. Rev. {\bfseries D83}
  (2011) 114039}
{\ttfamily [\href{http://arxiv.org/abs/1103.4835}{arXiv:1103.4835}]}.

\bibitem{Grinstein:2011dz}
B.~Grinstein, A.~L.~Kagan, J.~Zupan, and M.~Trott,
 \href{http://dx.doi.org/10.1007/JHEP10(2011)072}{JHEP {\bfseries 1110} (2011)
  072}
{\ttfamily [\href{http://arxiv.org/abs/1108.4027}{arXiv:1108.4027}]}.

\bibitem{Yan:2011tf}
K.~Yan, J.~Wang, D.~Y.~Shao, and C.~S.~Li,
 \href{http://dx.doi.org/10.1103/PhysRevD.85.034020}{Phys.Rev. {\bfseries D85}
  (2012) 034020}
{\ttfamily [\href{http://arxiv.org/abs/1110.6684}{arXiv:1110.6684}]}.

\end{thebibliography}\endgroup
\end{document}